\documentclass[journal]{IEEEtran}
\usepackage{cite}
\usepackage{amsmath,amssymb,amsfonts}
\usepackage{algorithmic}
\usepackage{graphicx,color}
\usepackage{textcomp}
\usepackage{subfigure}
\usepackage{comment}

\begin{document}

\title{Know What, Know Why: Semantic Hazard Communication for Intelligent V2X Systems}

\author{Chen Sun,~\IEEEmembership{Senior Member,~IEEE,}
        Wenqi Zhang,~\IEEEmembership{Member, IEEE,}\\
        Bizhu Wang,~\IEEEmembership{Member, IEEE,}
        Xiaodong Xu,~\IEEEmembership{Senior Member,~IEEE}\\
        Chau Yuen,~\IEEEmembership{Fellow,~IEEE,}
        Yan Zhang,~\IEEEmembership{Fellow,~IEEE,}
        and Ping Zhang,~\IEEEmembership{Fellow,~IEEE}
\thanks{C. Sun, W. Zhang are with Sony China Research Laboratory. B. Wang, X. Xu and P. Zhang are with Beijing University of Posts and Telecommunications, C. Yuen is with Nanyang Technological University. Y. Zhang is with University of Electronic Science and Technology of China.}
\thanks{Corresponding author: Wenqi Zhang, e-mail: wenqi.zhang@sony.com}
\thanks{Manuscript received xxx, 20xx; revised xxx, 20xx.}}

\markboth{Journal of \LaTeX\ Class Files,~Vol.~14, No.~8, August~2025}%
{Shell \MakeLowercase{\textit{et al.}}: Bare Demo of IEEEtran.cls for IEEE Journals}

\maketitle

\begin{abstract}
In current vehicle-to-everything (V2X) communication systems, roadside units (RSUs) broadcast brief warning messages that alert nearby vehicles to avoid potential hazards. However, these messages lack contextual information on why a warning is issued, leading to excessive caution or inefficient driving behaviors. To avoid such a situation, we propose a semantic-enhanced and explainable V2X (SEE-V2X) system. In the proposed system, RSUs equipped with smart cameras detect obstructions and transmit context-aware messages to vehicles. By understanding both what the hazard is and why it occurs, drivers can make more intelligent decisions based on their specific driving situation. Furthermore, through a real-field demonstration, we show the new ``see-through" feature in the proposed system, which enables drivers to visualize hidden pedestrians behind obstacles. We also perform simulations to compare traditional V2X with SEE-V2X under different traffic conditions. The results show that SEE-V2X significantly improves traffic efficiency and reduces unnecessary deceleration. 
\end{abstract}

\begin{IEEEkeywords} 
Vehicle-to-everything (V2X), semantic communication (SemCom), 3rd Generation Partnership Project (3GPP) standards.
\end{IEEEkeywords}

\IEEEpeerreviewmaketitle

\section{Introduction}
\IEEEPARstart{V}{ehicle}-to-everything (V2X) communications have become a fundamental component of intelligent transportation systems, enabling real-time exchange of critical safety information between vehicles, road infrastructure, pedestrians and cloud services. V2X is increasingly integrated into commercial vehicle platforms and regulatory frameworks, including their official inclusion in China’s 2024 New Car Assessment Program (C-NCAP). In particular, vehicle-to-vehicle (V2V), vehicle-to-infrastructure (V2I) and vehicle-to-pedestrian (V2P) are now considered essential features in next-generation vehicles.

In practice, V2X messages are transmitted through the cellular sidelink interface based on the 3rd generation partnership project (3GPP), also known as Cellular V2X (C-V2X), or through dedicated short-range communication (DSRC) based on IEEE 802.11p \cite{10056390}. These standards support direct, low-latency communication between user equipment (UE), such as vehicles and roadside units (RSUs), to broadcast messages about nearby hazards, including obstacles, collisions, or crossing pedestrians. Recently, V2X has been adopted in various commercial deployments and regulatory testbeds to enhance road safety. However, existing V2X warning systems are designed primarily to send short and context-free alert messages with limited bandwidth, especially when operating over 3GPP Release~15 sidelink channels \cite{5gaa2021c-v2x}. These messages indicate the presence of a hazard, without conveying specific information about its nature or underlying cause. For example, if a pedestrian is detected near an RSU, all nearby vehicles will receive an identical warning, regardless of their lane, direction, or likelihood of encountering the pedestrian. As a result, vehicles unnecessarily slow down or stop, even when they are not affected, leading to inefficiencies in traffic flow and potentially causing congestion over a wide area.

Semantic communications (SemCom) was introduced by Shannon in the 1960s as the second level of communications \cite{10183789}. Recent research activities have considered the use of SemCom to reduce latency and improve reliability in connected vehicles by compressing and transmitting sensor data \cite{10183789,9679803}. Performance in terms of bandwidth and computation delay is analyzed to justify the application of SemCom for V2X \cite{10574825}. In \cite{10695151}, SemCom is used to reduce the bandwidth required to share sensing information among multiple vehicles engaged in cooperative tasks. The authors in \cite{11048572} defined an importance score to quantify the semantic content of images and optimize image transmission accordingly in V2X networks. These previous studies have focused primarily on analyzing the performance of SemCom in V2X systems from a system-level perspective, evaluating metrics such as bandwidth efficiency and computational delay to justify its technical feasibility. However, the practical advantages of SemCom from the consumer's perspective, specifically regarding driving decision support, the interpretability of warnings, and overall traffic experience, remain largely underinvestigated.

To address the aforementioned limitations, this article proposed the sematic-enhanced and explainable V2X (SEE-V2X) system. Specifically, SemCom shifts the focus from the transmission of raw data or superficial alerts to the communication of high-level, context-sensitive meanings. This enables vehicles to understand the physical meaning of a warning. For instance, if a pedestrian is detected behind a parked vehicle, a SemCom-enabled RSU can extract the semantic representation of the scene (e.g., object type, position, relevance), encode the most important information, and transmit it to nearby vehicles. The receiving vehicle can then reconstruct the hazard using a semantic decoder and visualize it in real time, creating an augmented ``see-through" effect on the vehicle display. This enables the driver to assess the need for action based on the alignment of the lane, the direction of travel, and the predicted path. Furthermore, to validate this concept, we implemented a real-world prototype using a smart camera deployed on the RSU. The camera detects pedestrians and sends images to a semantic encoder running in the RSU. The encoded semantic information is transmitted via 5G network to a vehicle equipped with on-board semantic decoders. The vehicle then reconstructs the scene and overlays pedestrian information on a front-facing augmented reality (AR) display. For the purpose of comparison, a conventional V2X alert is displayed concurrently. The demonstration shows that the semantic overlay enables the driver to accurately assess the situation, avoiding unnecessary deceleration when the pedestrian is not in the his/her driving path. 

The rest of this article is organized as follows. Section II provides a comprehensive overview of the V2X communication protocol stack and current issues of the V2X system. The proposed SEE-V2X system is shown in Section III. Section IV introduces the field demonstration of the ``see-through" awareness.  Section V shows the evaluation of traffic efficiency. Section VI presents the next step of applying SemCom to V2X and the path to standardization ahead. Section VII concludes the article.

\section{Current V2X Deployment and Limitations}
The 2024 edition of the China New Car Assessment Program (C-NCAP) officially incorporates V2X capabilities into its vehicle safety assessment criteria. This policy has accelerated the adoption of C-V2X technology in both vehicles and road infrastructure. In response, multiple automotive manufacturers have announced that the upcoming vehicle models will include native support for C-V2X communication. The progress in the deployment of V2X is due to the rapid development of industrial standards. 

The V2X systems relied on lower layer protocols, such as the 3GPP sidelink and the IEEE DSRC technologies, that define the radio transmission, routing, and physical delivery of information. Meanwhile, message layer protocols, typically implemented at the application layer, define the syntax, structure, and semantics of V2X messages used by various safety-critical applications \cite{7992934}. 

Regarding the lower layer protocols, 3GPP introduced the V2X functionality in Release~14, which enables sidelink communication (PC5 interface) for basic safety services, including V2V, V2I, and V2P messages, all operating without reliance on the coverage of the cellular network. The enhancements in Release~14 included support for use cases such as platooning. Later, Release~16 of 3GPP introduced the 5G New Radio (NR) capabilities for V2X, which incorporated unicast, multicast and broadcast transmission modes. These updates supported advanced applications including cooperative driving, sensor sharing, and remote driving, offering improved sidelink reliability and reduced latency \cite{10078378}. Currently, chipsets designed based on Release~14 remains the dominant choice in commercial deployments, while NR-based solutions are still in early adoption stages.

As for the message layer protocols, the message formats are customized across different regions according to national standards and use-case requirements by European Telecommunications Standards Institute (ETSI), Society of Automotive Engineers (SAE) International and China Communications Standards Association (CCSA), etc. For example, in America, Basic Safety Messages (BSMs) provide essential information, such as vehicle position. In Europe, Cooperative Awareness Messages (CAMs) provide real-time awareness of surrounding entities. These messages are periodically broadcasted at 100$ms$ interval. Although many messages are in sync with SAE and ETSI, China has defined messages considering China traffic scenarios. The Chinese Roadside Information (RSI) message combines elements from the ETSI Decentralized Environmental Notification Message (DENM) and the In-Vehicle Information (IVI), while extending them to China-specific use cases. These include road hazard warnings, traffic control instructions, driving instructions, and static infrastructure information. These enhancements cover a wide range of scenarios, including road hazard alerts, dynamic traffic control information, driving guidance, and static infrastructure descriptions.

To evaluate the current performance of V2X, we performed a test at a T-junction in Wuhan Economic Development District \cite{Cui22}. In this controlled environment, 200 RSUs were deployed along the road to emulate a dense C-V2X traffic scenario. The results showed a stable packet reception rate exceeding $95\%$ within a range of $50m$, indicating high communication reliability under typical operating conditions. However, since traditional V2X messages lack semantic context, all receiving vehicles react to a broadcast warning in the same way, typically by slowing down or stopping, even if the hazard is not directly relevant to their path. This lack of contextual awareness contributes to unnecessary congestion and degraded traffic efficiency in a broad area.
\begin{figure*}[h]
	\centering 
    \label{fig.sys}
	\subfigbottomskip=0.5cm 
	\subfigcapskip=-5pt 
	\subfigure[Straight urban expressway scenario]{
		\includegraphics[width=0.95\linewidth]{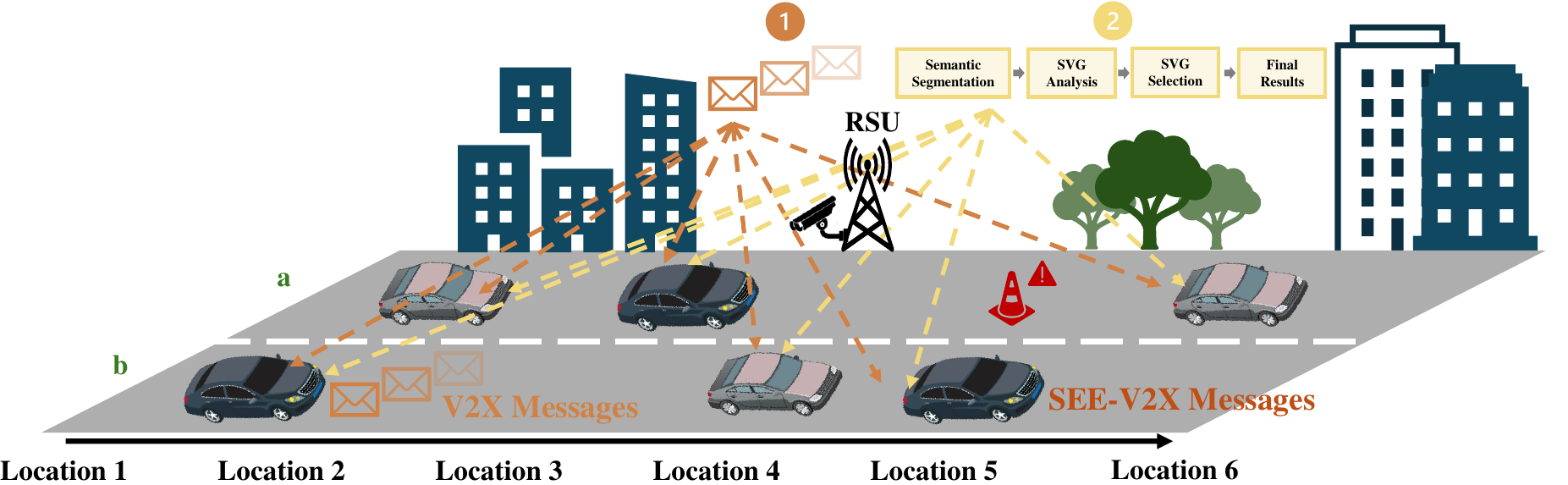}}
	  \\
	\subfigure[Intersection scenario]{
		\includegraphics[width=0.95\linewidth]{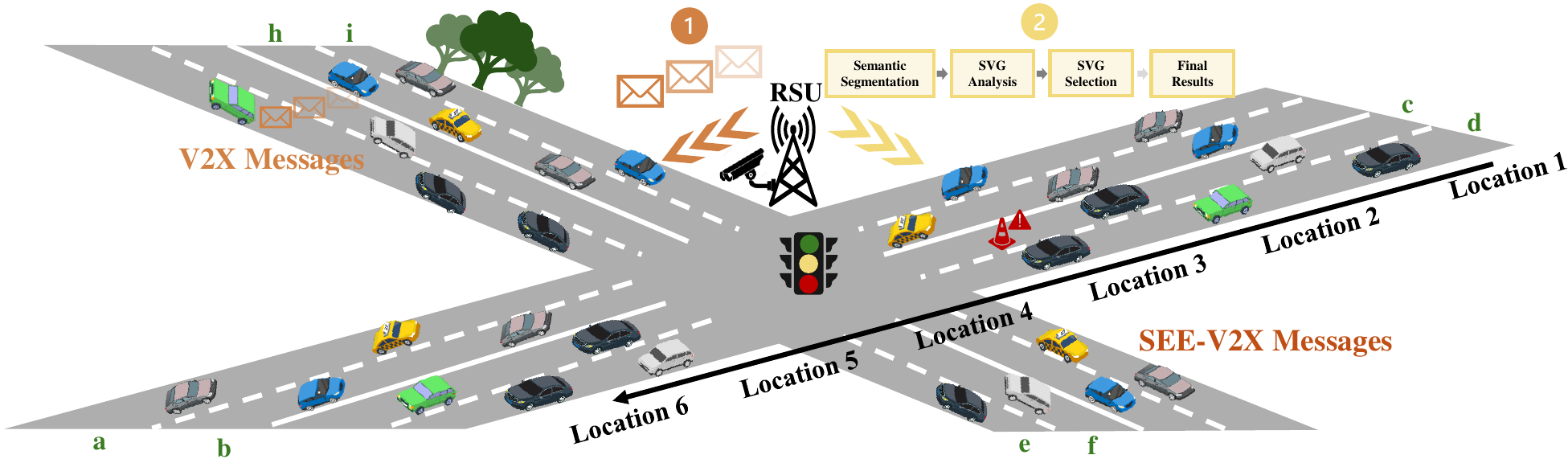}}
	\caption{System architecture showing the possible locations of RSU and the flows of 1) traditional V2X message from smart camera to vehicle via RSU in V2X band; 2) semantic information from smart camera, semantic encoder and RSU to semantic decoder with vehicle display through 5G link.}
\end{figure*}

\section{How SEE-V2X Works}
We design and implement the SEE-V2X system that enables intelligent hazard perception and reasoning. Fig.~1 shows the architectural design. For traditional V2X a smart camera installed on an RSU broadcasts messages to vehicles. For SEE-V2X, a smart camera, along with a semantic encoder, is integrated into an RSU, while a semantic decoder is located within the vehicle. We consider both straight urban highway and intersection scenarios in which vehicles receive traditional V2X or SEE-V2X warnings from RSUs deployed in different possible locations. The smart camera is equipped with integrated artificial intelligence (AI) capabilities to detect objects in real time, identifying pedestrians in the scene. Upon detecting a pedestrian, those obstructed by nearby obstacles such as parked vehicles, the camera triggers two processes in parallel: 1) broadcasting a conventional V2X alarm message via the RSU to vehicles through the sidelink, and 2) sending the captured image to a semantic encoder for further processing, which subsequently sends semantic information via RSU to the decoder on vehicle through the 5G network. 

\subsection{Semantic Encoding and Transmission}
The semantic encoder processes the input image using a scene graph generation (SGG) model with nonlinear transform source-channel coding (NTSCC) \cite{9791398}. The SGG model identifies key entities and relationships within the scene, and the NTSCC encodes them into semantic information, prioritized according to importance to vehicular safety. For example, we can assign the highest priority to the ``human" class. The semantic encoder, which is deployed on the edge RSU in production deployments, operates on a high performance computing platform (NVIDIA RTX 4090 GPU). The encoded semantic information is transmitted to the vehicle using a 5G network link. The encoding process is adaptive to the quality of the wireless link and selects only the most essential objects (such as roads and pedestrians) based on available bandwidth and latency requirements. The 5G network provides an interface between the core network and the application so that the application can query and demand the quality of the wireless link, for example, through the CAMARA application programming interface (API)\cite{CAMARA2025}.

\subsection{Semantic Decoding and Augmented Display}
On the vehicle side, a semantic decoder receives the transmitted semantic information and reconstructs the high-level semantic representation of the detected objects. The pedestrian's position can be aligned with the vehicle's camera view onboard based on the estimated location of the pedestrian provided by the smart camera. This enables the reconstructed pedestrian to be overlayed onto the live video fed from the vehicle’s front-facing camera, creating a real-time AR effect. This visualization enables the driver (or autonomous controller) to ``see through” occlusions and understand the spatial cause of the warning.
\begin{figure*}[h]
    \centering
    \includegraphics[width=1\linewidth]{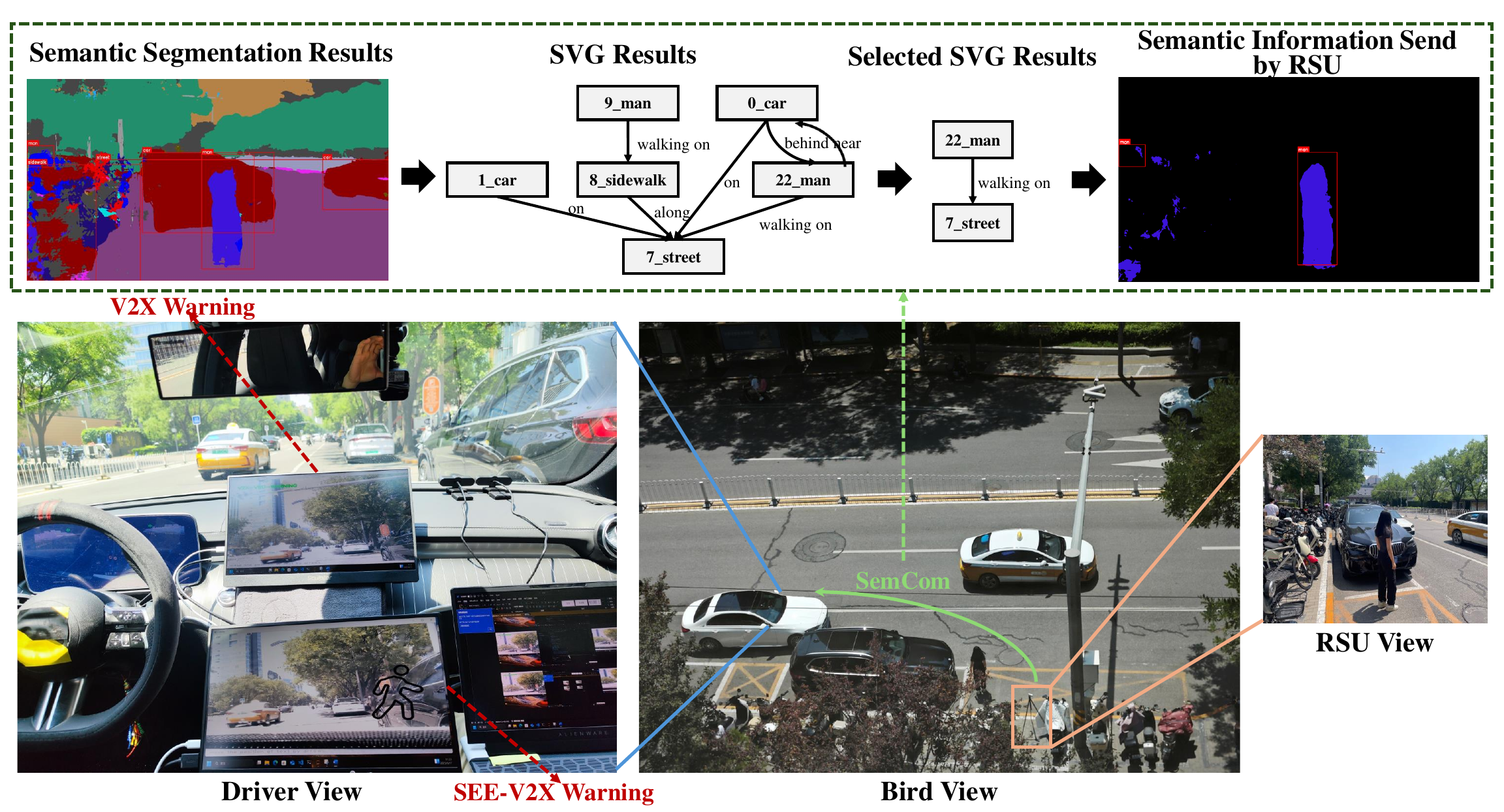}
    \caption{Field demonstration of traditional V2X warning (showing a text message on the left upper corner of the screen) and SEE-V2X warning with ``see-through" effect (showing a transparent human overlaid on real scenery).}
    \label{Fig.demo}
\end{figure*}

\vspace{-0.3cm}
\subsection{Training and Deployment Considerations}
The semantic encoder and decoder are jointly trained using Cityscapes and COCO datasets that include annotated urban scenes with pedestrians and occlusions. The model learns to prioritize and reconstruct safety-critical information under varying communication constraints. Compared to traditional V2X systems, the proposed system requires AI capabilities on both the transmitter (i.e., RSU/camera) and receiver (vehicle) sides. In addition, the proposed system illustrates the fundamental shift from data-level alerts to semantically meaningful communication, enabling vehicles to receive hazard warnings and understand the causes behind them, leading to safer and more context-aware driving behavior.

\section{Field Demo: ``See-Through” Awareness}
As shown in Fig.~2 the field demonstration is conducted on a public street in Beijing, where vehicles are parked on the street side. Specifically, this is a typical urban scenario in which pedestrians may emerge unexpectedly from behind vehicles. Such occlusions present a substantial safety risk, particularly when drivers lack direct visual access to the crossing area. To address this challenge, a smart camera is deployed on the RSU infrastructure. In particular, the smart camera features integrated edge AI processing, enabling real-time pedestrian detection. The embedded model is specifically trained to identify human figures with high precision. Upon detecting a pedestrian, the smart camera issues a hazard notification to the RSU, which subsequently broadcasts a standard V2X alert to surrounding vehicles. At the same time, the semantic encoder extracts high-level semantic information related to the pedestrian and transmits the encoded semantic information via the 5G network to a processing unit located inside the vehicle. On the vehicle side, a semantic decoder reconstructs the visual representation of the pedestrian from the semantic information received. Furthermore, we employ CAMARA to emulate the 5G core network that provides API for network quality query. In particular, the quality of the network is divided into three levels. Based on the quality of the network, the semantic information corresponding to the most important objects are transmitted accordingly. 

The reconstructed image is then overlayed onto a live video stream captured from a dashboard-mounted display device positioned at the front of the vehicle cockpit. This enhancement creates the ``see-through" effect, allowing the driver to visualize the occluded pedestrian in real time as if the obstacle (e.g., a parked vehicle) were transparent. The field demonstration validates the feasibility of SEE-V2X communication to deliver not only hazard alerts, but also visual context, significantly improving driver awareness and safety. The system enables the driver to understand the cause of the alert. A decision to reduce speed can be made based on the assessment of whether the pedestrian constitutes a threat to the vehicle.
\begin{figure*}[h]
\label{fig.Simu}
\centering
\subfigure[]{
\begin{minipage}[b]{0.49\textwidth}
\includegraphics[width=1\textwidth]{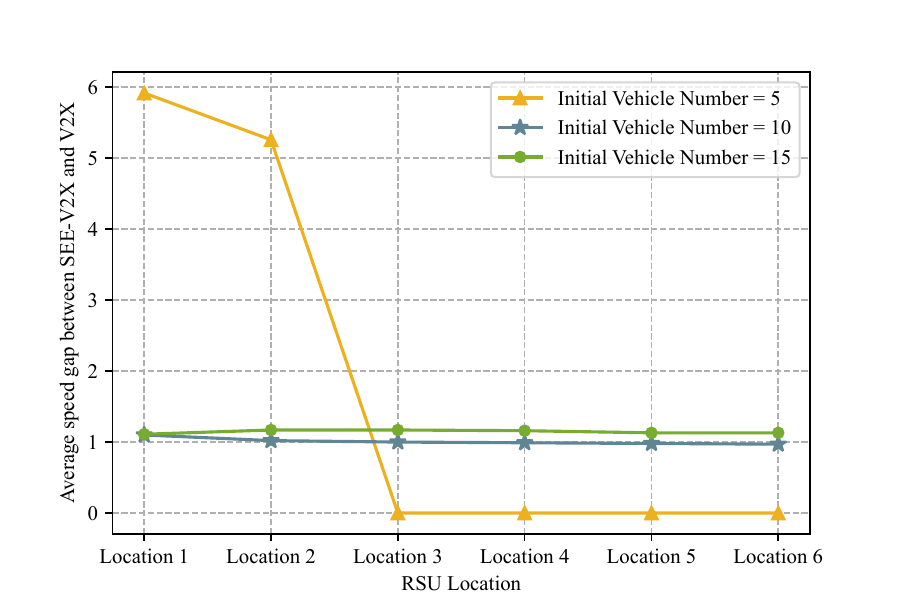}
\end{minipage}}
\subfigure[]{
\begin{minipage}[b]{0.49\textwidth}
\includegraphics[width=1\textwidth]{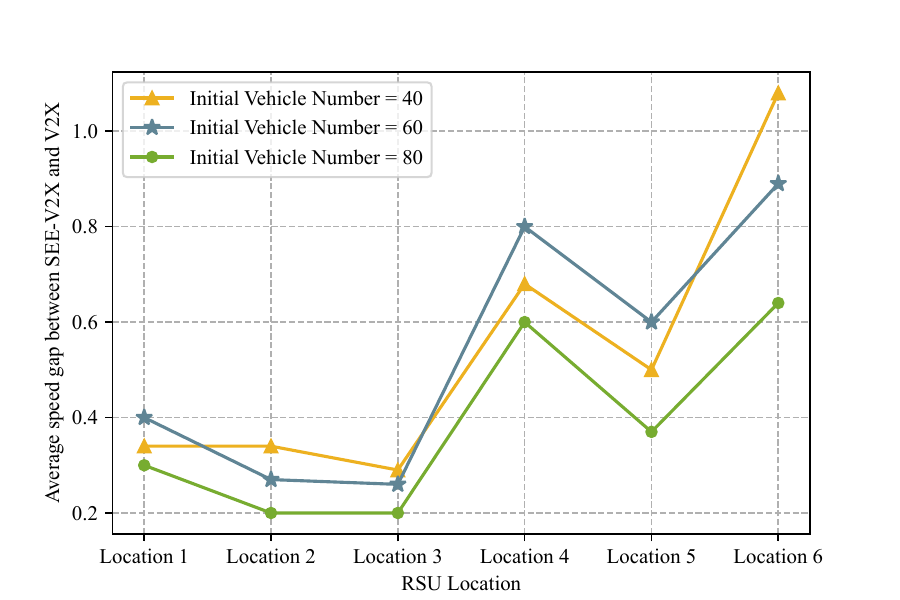} 
\end{minipage}}
\caption{Impact of RSU location on speed gap [$m/s$] between SEE-V2X and traditional V2X in both (a) straight urban expressway scenario and (b) intersection scenario. (For the two scenarios, the initial speed of vehicles is set to 20$m/s$, and the RSU locations are indicated as Fig. 1.)}
\end{figure*}
\section{Simulation: Measuring Traffic Efficiency}
To evaluate the benefits of transportation efficiency of the proposed SEE-V2X system compared to conventional V2X frameworks, we develop a simulator based on the Intelligent Driver Model (IDM) \cite{10288528}. This simulator enables a comparative analysis of traffic flow improvements in various traffic scenarios. Detailed performance results are presented in Fig.~3. Specifically, in the scenario of the straight urban expressway, the performance benefits of the proposed SEE-V2X system become apparent under low vehicle density conditions when the RSU is positioned at location~1 and location~2 near the urban expressway entry point. In this area, vehicles actively adjust inter-spacing to establish safe following distances. The slowdown of one car often leads to a transient clustering of vehicles. The SEE-V2X facilitates more intelligent coordination during this critical phase by allowing vehicles to interpret contextual hazard information, thus minimizing unnecessary deceleration and enhancing traffic smoothness. In contrast, the traditional V2X system, which lacks this contextual awareness, tends to trigger cautious behavior across all receiving vehicles regardless of relevance, leading to inefficiencies. However, when the RSU is deployed further downstream in the region where the vehicles have already stabilized their spacing, the incremental benefit of SEE-V2X decreases. In these cases, minor decelerations induced by hazard alerts have limited ripple effects on the broader traffic flow due to the naturally dispersed and steady vehicle distribution.

In the intersection scenario, vehicle density is generally higher than in a straight expressway, resulting in more complex interactions. As the RSU is progressively placed closer to the center of the intersection, the effectiveness of the proposed SEE-V2X system becomes increasingly apparent. When the RSU is deployed at the entering area, i.e., location~1 and location~2, there are only a few vehicles that start to enter intersection at lane~c and lane~d. The V2X messages reach a limited number of vehicles. Thus, the benefit of SEE-V2X over traditional V2X is not significant. When approaching the intersection, vehicles dynamically adjust their speed in response to traffic conditions at the intersection. The RSU at location~3 detects obstacles and sends messages covering more vehicles. The SEE-V2X enables more precise coordination by conveying context-specific hazard information. This leads to fewer unnecessary decelerations compared to traditional V2X, where alerts are broadcast without relevance filtering. However, once the RSU is at location~5, the intersection center, the marginal benefits of SEE-V2X diminish. At this point, vehicle behavior is predominantly governed by the inherent requirement to yield or stop due to cross-traffic restrictions. This limits the influence of any V2X mechanism. Beyond the intersection at location~6, as vehicles transition back into a more stable formation with adequate separation, the advantages of SEE-V2X reemerge. During this phase, context awareness allows for more targeted responses to localized hazards, thus enhancing traffic throughput and reducing congestion more effectively than the traditional V2X system.

\section{Industrial Challenges and Standards Landscape}
\subsection{Challenges}
SemCom empowers the transmission of context-aware information that captures the intent, location, and criticality of the hazard. This enables each vehicle to interpret the situation based on its own trajectory, lane, and proximity to the hazard. Vehicles equipped with SEE-V2X can make more precise decisions by replacing rigid one-size-fits-all warnings with semantically rich messages. This approach reduces unnecessary reactions and improves both safety and traffic efficiency. To fully understand these benefits, we built a real-world demonstration of SemCom for V2X applications and identified the following challenges.

\subsubsection{Edge-Compute Cost and 5G Coverage Gaps}
The advent of robust on-device AI marks a notable transition from an exclusive dependence on cloud-based solutions. This transition to the terminal side is driven by a dramatic intersection of two exponential trends. On the other hand, Moore's law has historically seen computing power double roughly every 18 months. On the other hand, the density of large language models (LLMs) is growing at a staggering rate, doubling approximately every three months \cite{Liuzhiyuan}. The convergence of these two growth curves in 2024 signifies a pivotal moment: terminal devices now possess the requisite power to run increasingly sophisticated AI models locally. This allows terminals to implement a semantic encoder or decoder. However, to support real-time applications, such as remote driving, the system should achieve inference speeds that align with typical camera frame rates (e.g., 30 frames per second). Future efforts will focus on reducing model complexity through techniques such as knowledge distillation using domain-specific traffic datasets, enabling lightweight deployment on edge or onboard units.    

\subsubsection{Knowledge Mismatch Risks and Model Updates}
SemCom inherently depends on shared knowledge between the semantic encoder (e.g., RSU-side perception model) and the decoder (e.g., vehicle-side interpretation model). However, vehicles often traverse unfamiliar environments where the decoder-side models may not be aligned with the local semantics produced by roadside encoders. This knowledge mismatch leads to degraded interpretation of semantic messages. Future work will explore domain adaptation, transfer learning, and federated learning of decoder models to enable SEE-V2X in diverse scenarios.

\subsubsection{Protocol and Architecture Integration}
Our demonstration is implemented in the application layer. The semantic message from the encoder is sent as a payload on TCP/IP protocols. This is constrained by legacy communication module designs based on layered protocol stacks. For SemCom to achieve its full potential, joint optimization of AI model inference and physical layer transmission is essential. This vision calls for a rethinking of communication architectures, including cross-layer protocol co-design, hardware-software co-optimization, and the development of next-generation communication chipsets. These advances are expected to materialize within the 6G system roadmap by 2030.

\subsubsection{Scalability and Bandwidth Efficiency}
Although SemCom significantly reduces bandwidth usage compared to raw sensor data transmission, message volume still scales with application penetration and vehicle density. Efficient semantic encoding strategies should adaptively balance information richness, task relevance, and channel capacity. Dynamic encoding schemes that consider both application priorities and real-time network status will be critical in dense urban environments.

\subsubsection{Extending Semantic Content and Applications}
Future extensions may incorporate richer environmental semantics, including moving objects, road conditions, and traffic flow dynamics, thereby improving the granularity of decision making. Additionally, AI-generated content (AIGC) techniques can be integrated with semantic decoding to improve driver interfaces, such as rendering hazard visualizations in different styles to increase engagement or emotional significance. Proactive safety mechanisms can be further improved by incorporating vehicle-side prediction models capable of reasoning about semantic cues and forecasting potential risks.

\subsubsection{Security and Privacy}
As semantic messages carry high-level perceptual insights, ensuring their integrity, authenticity, and confidentiality will be crucial. Future research should address the challenges of secure semantic encoding, tamper detection, and privacy-preserving data sharing, especially as SemCom becomes integral to real-world V2X deployments.

\subsection{Standards Landscape} 
In a recent meeting of the China Intelligence Transportation System Industry Alliance (C-ITS) led by the Research Institute of Highway Ministry of Transport (RIOH), the enhanced V2X message layer has been proposed to include the descriptive messages of the scene as an extension of the V2X messages. By defining the messages and grammar, the extended messages gives a language description of the scenario at the expense of radio resources. The autonomous driving entities receiving these messages are supposed to have a certain reaction defined by the standards. Unlike sharing sensing data such as image, point cloud data, etc. of the scene, this approach can be considered as sharing the description text of the scene. However, this approach is still different from the SemCom whose essence is enabling the receiver to infer the intended meaning from information sent by the transmitter based on the alignment of knowledge at both sides.

Along with the advancement of AI technologies, the SemCom has become a hot topic in academia. In March 2025, 3GPP held the 6G workshop in Inchon, South Korea \cite{3GPP6G}. In this meeting, various companies presented their views on the direction of the 6G technology. It was the first time SemCom was considered as one of the candidates for 6G communication. In this meeting, Huawei and LGE presented the industry envisions a paradigm shift from traditional bit-centric transmission to communication that prioritizes meaning and intent, which is proposed to facilitate interaction among AI agents and support embodied/spatial AI services. Lenovo suggested that the new architecture uses AI at both the transmitter and the receiver to jointly optimize the source and channel coding to be evaluated against traditional separate source and channel coding schemes. This approach promises significant gains in efficiency and reliability, notably through reduced data overhead, decreased network load, and improved energy efficiency. This necessitates the definition of new KPIs tailored for SemCom as proposed by VIAVI. These efforts will push the development of industrial standards moving forward towards realization of SemCon in the near future.
    
\section{Conclusion}
The proposed SEE-V2X system marks a critical step toward AI-native transportation systems, enabling communication that not only informs the presence of hazards, but also explains the reasons behind them. By transmitting distilled, context-aware information from roadside perception units to vehicles, the proposed system improves driving safety and overall traffic efficiency. Through both field demonstration and simulation studies, we demonstrated that SEE-V2X outperforms traditional V2X messaging in scenarios where the direct line of sight is obstructed and enables drivers to make more informed and context-sensitive decisions. In particular, drawing on insights from field deployments, we identified a fundamental limitation of current V2X systems. In summary, this work establishes a foundation for intelligent V2X systems enabled by SemCom, offering a pathway to safer, smarter, and more context-aware cooperative driving. Continued advances in model optimization, system integration, and adaptive communication will be essential to realize the full vision of AI-native vehicular networks.

\ifCLASSOPTIONcaptionsoff
  \newpage
\fi

\bibliographystyle{IEEEtran}
\bibliography{ref}

\begin{thebibliography}{10}
\providecommand{\url}[1]{#1}
\csname url@samestyle\endcsname
\providecommand{\newblock}{\relax}
\providecommand{\bibinfo}[2]{#2}
\providecommand{\BIBentrySTDinterwordspacing}{\spaceskip=0pt\relax}
\providecommand{\BIBentryALTinterwordstretchfactor}{4}
\providecommand{\BIBentryALTinterwordspacing}{\spaceskip=\fontdimen2\font plus
\BIBentryALTinterwordstretchfactor\fontdimen3\font minus \fontdimen4\font\relax}
\providecommand{\BIBforeignlanguage}[2]{{%
\expandafter\ifx\csname l@#1\endcsname\relax
\typeout{** WARNING: IEEEtran.bst: No hyphenation pattern has been}%
\typeout{** loaded for the language `#1'. Using the pattern for}%
\typeout{** the default language instead.}%
\else
\language=\csname l@#1\endcsname
\fi
#2}}
\providecommand{\BIBdecl}{\relax}
\BIBdecl

\bibitem{10056390}
E.~Moradi-Pari, D.~Tian, M.~Bahramgiri, S.~Rajab, and S.~Bai, ``{DSRC} versus {LTE-V2X}: Empirical performance analysis of direct vehicular communication technologies,'' \emph{IEEE Trans.on Intelligent Transportation Systems}, vol.~24, no.~5, pp. 4889--4903, 2023, doi={10.1109/TITS.2023.3247339}.

\bibitem{5gaa2021c-v2x}
\BIBentryALTinterwordspacing
5GAA, ``List of {C-V2X} devices,'' {5G} Automotive Association, Tech. Rep., November 2021. [Online]. Available: \url{https://5gaa.org/list-of-c-v2x-devices/}
\BIBentrySTDinterwordspacing

\bibitem{10183789}
Y.~Shi, Y.~Zhou, D.~Wen, Y.~Wu, C.~Jiang, and K.~B. Letaief, ``Task-oriented communications for {6G}: Vision, principles, and technologies,'' \emph{IEEE Wireless Commun.}, vol.~30, no.~3, pp. 78--85, 2023, doi={10.1109/MWC.002.2200468}.

\bibitem{9679803}
X.~Luo, H.-H. Chen, and Q.~Guo, ``Semantic communications: Overview, open issues, and future research directions,'' \emph{IEEE Wireless Commun.}, vol.~29, no.~1, pp. 210--219, 2022, doi={10.1109/MWC.101.2100269}.

\bibitem{10574825}
J.~M. Gimenez-Guzman, I.~Leyva-Mayorga, and P.~Popovski, ``Semantic {V2X} communications for image transmission in {6G} systems,'' \emph{IEEE Network}, vol.~38, no.~6, pp. 48--54, 2024, doi={10.1109/MNET.2024.3420214}.

\bibitem{10695151}
Y.~Feng, H.~Shen, Z.~Shan, Q.~Yang, and X.~Shi, ``Semantic communication for edge intelligence enabled autonomous driving system,'' \emph{IEEE Network}, vol.~39, no.~2, pp. 149--157, 2025, doi={10.1109/MNET.2024.3468328}.

\bibitem{11048572}
A.~Cai, L.~Wang, Y.~Lin, C.~Liu, and P.~Qian, ``Semantic importance-aware image transmission in {V2X} networks,'' \emph{IEEE Internet of Things J.}, pp. 1--1, 2025, doi={10.1109/JIOT.2025.3582442}.

\bibitem{7992934}
S.~Chen, J.~Hu, Y.~Shi, Y.~Peng, J.~Fang, R.~Zhao, and L.~Zhao, ``Vehicle-to-everything ({V2X}) services supported by {LTE}-based systems and {5G},'' \emph{IEEE Commun. Stand. Mag.}, vol.~1, no.~2, pp. 70--76, 2017, doi={10.1109/MCOMSTD.2017.1700015}.

\bibitem{10078378}
A.~Bazzi, C.~Campolo, V.~Todisco, S.~Bartoletti, N.~Decarli, A.~Molinaro, A.~O. Berthet, and R.~A. Stirling-Gallacher, ``Toward {6G} vehicle-to-everything sidelink: Nonorthogonal multiple access in the autonomous mode,'' \emph{IEEE Veh. Technol. Mag.}, vol.~18, no.~2, pp. 50--59, 2023, doi={10.1109/MVT.2023.3252278}.

\bibitem{Cui22}
\BIBentryALTinterwordspacing
T.~Cui, L.~Li, Z.~Zhang, and C.~Sun, ``{C-V2X} vision in the chinese roadmap: Standardization, field tests, and industrialization,'' in \emph{Vehicular Networks}, A.~Haidine, Ed.\hskip 1em plus 0.5em minus 0.4em\relax Rijeka: IntechOpen, 2022, ch.~2, doi = {10.5772/intechopen.107933}. [Online]. Available: \url{https://doi.org/10.5772/intechopen.107933}
\BIBentrySTDinterwordspacing

\bibitem{9791398}
J.~Dai, S.~Wang, K.~Tan, Z.~Si, X.~Qin, K.~Niu, and P.~Zhang, ``Nonlinear transform source-channel coding for semantic communications,'' \emph{IEEE J. Sel. Areas in Commun.}, vol.~40, no.~8, pp. 2300--2316, 2022, doi={10.1109/JSAC.2022.3180802}.

\bibitem{CAMARA2025}
{CAMARA Project}, \url{https://camaraproject.org/}.

\bibitem{10288528}
S.~Fang, L.~Yang, X.~Zhao, W.~Wang, Z.~Xu, G.~Wu, Y.~Liu, and X.~Qu, ``A dynamic transformation car-following model for the prediction of the traffic flow oscillation,'' \emph{IEEE Intell. Transp. Syst. Mag.}, vol.~16, no.~1, pp. 174--198, 2024, doi={10.1109/MITS.2023.3317081}.

\bibitem{Liuzhiyuan}
C.~Xiao, J.~Cai, W.~Zhao, G.~Zeng, B.~Lin, J.~Zhou, Z.~Zheng, X.~Han, Z.~Liu, and M.~Sun, ``Densing law of {LLMs},'' 2024. [Online]. Available: https://arxiv.org/abs/2412.04315.

\bibitem{3GPP6G}
\BIBentryALTinterwordspacing
``{3GPP} workshop on {6G},'' Inchon, South Korea, 2025. [Online]. Available: \url{https://www.3gpp.org/news-events/3gpp-news/6gworkshop-2025}
\BIBentrySTDinterwordspacing

\end{thebibliography}

\end{document}